\documentclass[letterpaper, 10 pt, conference]{ieeeconf}
\usepackage{amsmath,amsfonts}
\usepackage{algorithmic}
\usepackage{algorithm}
\usepackage{array}
\usepackage[caption=false,font=normalsize,labelfont=sf,textfont=sf]{subfig}
\usepackage{textcomp}
\usepackage{stfloats}
\usepackage{url}
\usepackage{verbatim}
\usepackage{graphicx}
\usepackage{cite}
\usepackage{color}
\usepackage[nocomma]{optidef}
\DeclareMathOperator{\isgoal}{isgoal}
\DeclareMathOperator{\RDD}{RDD}
\DeclareMathOperator{\HR}{HR}
\DeclareMathOperator{\VR}{VR}
\DeclareMathOperator{\DF}{DF}
\DeclareMathOperator{\IF}{IF}

\IEEEoverridecommandlockouts
\begin{document}

\title{Respiratory-Aware Routing for Active Commuters}

\author{Abigail Langbridge, Pietro Ferraro and Robert Shorten
\thanks{A. Langbridge, R. Shorten and P. Ferraro are with the Dyson School of Design Engineering, Imperial College London.}}

\maketitle

\parindent2mm
\begin{abstract}
    Cyclists travelling in urban areas are particularly at risk of harm from particulate emissions due to their increased breathing rate and proximity to vehicles. In this paper we combine human respiratory models with models of particulate inhalation to estimate the pollution risk an individual is experiencing in real time given the local pollution level and their heart rate for the first time. Using this model as a baseline, we learn a policy that simultaneously optimises the route for a large number of cyclists with diverse origins and destinations, to minimise overall pollution risk and account for the detrimental impacts of congestion. We learn this policy using reinforcement learning techniques on simulated data in different environments with varying distributions of cyclist fitness. These findings establish that individualised routing is effective in reducing pollution risk while cycling, improving the net benefits of active commuting.
\end{abstract}

\vspace{0em}
\section*{Nomenclature}
\begin{table}[H]
\centering
\begin{tabular}{p{0.1\columnwidth} | p{0.75\columnwidth}}
\hline
\hline
Term & Definition\\
\hline
PM2.5 & Particulate matter with diameter less than 2.5 $\mu m$\\
RDD & Received deposition dose, the mass of particulates deposited in the lungs\\
MMD & Mass-median diameter, the diameter of the median-weighted particle in a given sample or mass fraction\\
VR & Ventilation rate, the volume of air breathed per minute\\
HR & Heart rate, the number of beats per minute\\
$P$ & Mechanical power required to traverse a segment\\
$\eta_{mech}$ & Mechanical efficiency of the bicycle transmission\\
$C_r$ & Resistance coefficient of rubber tyres on asphalt\\
$C_d$ & Drag coefficient of the cyclist\\
$m$ & Mass of the cyclist, bike and any luggage\\
$g$ & Gravitational constant\\
$\Delta h$ & Change in vertical height over a road segment\\
$l$ & Length of the road segment\\
$v$ & Travel velocity of the cyclist\\
$\rho$ & Air density\\
$A$ & An estimate of the frontal area of cyclist and bike\\
$t$ & Time taken to traverse the road segment\\
$\tau_r$ & Time coefficient for HR with step change in P\\
$c$ & Gradient defining steady-state HR with increasing P\\
\hline
\hline
\end{tabular}
\label{tab:nomencl}
\end{table}

\section{Introduction}
In 2018, 417,000 deaths were attributed to PM2.5 exposure across Europe\cite{PM_B_06}. These fine pollutant particles can penetrate far into the respiratory system when inhaled, leading to an increased occurrence of cardiac arrest, stroke, asthma, reduced lung function and dementia from both long and short-term exposures\cite{ShortTerm_PM_01}. Road vehicles are a significant contributor to PM2.5 concentrations in cities through emissions from incomplete combustion and tyre wear, with commuters receiving a high proportion of their daily exposure while commuting\cite{PM_B_01}. However, commute exposure can be difficult to quantify, resulting in limited awareness amongst cyclists of the potential risks of frequent or prolonged exposure while commuting\cite{PM_B_04}. Note, in contract to the prevailing narrative, vehicle electrification will not eliminate vehicle induced particulate generation as non-tailpipe emissions (tyre, road and brake emissions) are a growing problem due to the increased vehicle weight and power of electric vehicles.

With urban traffic forecast to increase 51\% by 2050\cite{AllAboard} and the cost of poor air quality to the NHS set to rise to £10 billion by 2035\cite{NHS_PMCost}, quantifying and mitigating the pollution risk of urban travel is critical. Interventions aimed at commuting are uniquely positioned to have huge impact on an individual's annual exposure, with the repetition of journeys providing an ideal environment for routing-based optimisation at an individual level.

Accordingly, in this paper we propose a personalised routing algorithm that provides paths to cyclists with the aim of minimising the effects of pollution. Our algorithm takes into account real-time measurements, such as the cyclist's heart rate and their physiological parameters. Due to the intrinsic complexity of the environment under consideration we make use of  Reinforcement Learning (RL) methods to train our controller on simulated data, in a variety of virtual environments to take into account varying level of distributions of cyclist fitness.

\section{Background}

In this paper we introduce the problem of pollution risk inequality between passive and active modes of transport, the problem of pollution-aware routing, and our use of reinforcement learning as our algorithm of choice.

\subsection{Pollution Risk Inequality}
There is a significant volume of work attempting to quantify personal PM2.5 exposure while commuting with various transport modes, both active and passive \cite{PM_B_04}\cite{PM_B_02}\cite{PM_B_03}. These studies report similar exposure trends, with walking and cycle journeys having the lowest exposure concentration per cubic metre of air compared to in-vehicle journeys. However, as active journeys can take much longer than passive ones and result in higher breathing rates, exposure is an insufficient metric to model the pollution risk of these modes.

It follows from these raised breathing rates that the total volume of particulate pollutants inhaled by an individual is also higher across an identical route for active transport\cite{PM_B_05}. To the best of the authors' knowledge, this is not considered in any previous work. This higher risk can be quantified using the received deposition dose (RDD, $\mu g$), which represents the mass of pollutant particles which remain in the lungs of an individual after exhalation. For an average adult male, RDD is reported to be more than 2.5 times higher for cycle than car journeys\cite{PM_B_01}, however little research has been conducted into the extent to which an individual's physiological factors can impact the validity of these generalised predictions.

\subsection{Reinforcement Learning}
RL is concerned with decision making in Markov Decision Processes (MDPs). MDPs are characterised by:

\begin{itemize}
    \item A state space $S$
    \item An action space $A$
    \item A reward function $R:S\times A \rightarrow \mathbb{R}$
    \item A discount factor $\gamma \in [0,1)$
    \item A transition probability $P:  S^2\times A \rightarrow [0,1]$
    \item A probability density $\pi:S \times A \rightarrow [0,1]$.
\end{itemize}

In this framework the objective of RL is to maximize the discounted total reward over an \emph{episode} of length $T$. Formally, let $\tau$ be a sequence $(s_t, a_t, r_t)$ with $a_t \sim \pi(\cdot|s_t)$, $s_{t+1} 
\sim P(\cdot, s_t, a_t)$ and $r_t$ being the reward at time $t$. The optimal policy $\pi^*$ is specified by:

\begin{equation}
    \pi^* = \arg\max \mathbb{E}_{\tau \sim \pi}\left[\sum_{t=1}^{T} \gamma^{t-1}r_t\right]
\end{equation}

One common technique to find $\pi^*$ is Proximal Policy Optimization (PPO) \cite{PPO}, an on-policy policy gradient algorithm. Given a $\theta-$parametrized policy, $\pi_\theta$, and a set of samples collected with it, PPO updates the parameters $\theta$ in order to maximise 
\begin{equation}
J^{PPO}(\theta) = \mathbb{E}_t \left[ \min\left( r_t(\theta)\tilde{A}_t, \; \text{clip}(r_t(\theta), 1-\epsilon, 1+\epsilon) \tilde{A}_t \right)\right],\nonumber
\end{equation}

where $\tilde{A}_t$ is an estimate of the advantage function and $r_t(\theta) = \dfrac{\pi_\theta (a|s)}{\pi_{\theta_{\text{old}}}(a|s)}$. The interested reader can find more details about the original implementation\footnote{Refer to \url{https://github.com/ray-project/ray/blob/master/rllib/algorithms/ppo/ppo.py} for the implementation used in this paper.} of PPO in \cite{PPO}.


\section{Pollution Risk Modelling}
Most existing literature focuses on two key factors affecting the pollution-routing problem: the level of pollution on each route segment and the duration of exposure. Works that do quantify pollution risk rely on generic models for estimating ventilation rates, and therefore the inhalation of pollution. A key contribution of this work is the development of a personalised model for estimating the pollution risk of active journeys, which additionally considers individuals' elevated breathing rates when modelling pollution risk. This model can be evaluated in real-time using heart rate and exposure data to produce a highly personalised estimate of pollution risk.

We note that the feasibility of collecting cheap and accurate real-time pollution data has been demonstrated in a number of projects; see, for example, the iSCAPE citizen sensing project which deployed a large-scale, low-cost fixed air quality monitoring system\cite{PM_04}. Further, with the continued growth in adoption of smart watches, heart rate data is widely collected and commonly provided to fitness and health tracking systems \cite{ni2019modeling}. 

The synthesis of this model also facilitates the calculation of individually pollution-optimal routes to be calculated using some dynamic programming approach, allowing individuals to make safe choices accounting for the pollution risk of a journey. Further, we suggest that these optimal routes will vary with physiological characteristics, resulting in fitness-dependent optimal routes.

\subsection{Ventilation Rate Modelling}
Ventilation Rate (VR) can be measured directly during exercise using spirometry equipment, however this is bulky and intrusive. Heart rate has been shown to be a good predictor of ventilation rate and is commonly used as a surrogate measurment. Zuurbier et al. present a sex-stratified log-linear model, Equation \ref{eq:vr}, derived from bicyle ergometer tests for use extimating VR in the TRAVEL study \cite{O_01}.

\begin{equation}
    VR = 
    \begin{cases}
    \text{\bf IF $\bf sex = M$ } e^{0.021 \cdot \HR + 1.03} \\
    \text{\bf ELSE } e^{0.023 \cdot \HR + 0.57} \\
    \end{cases}\\
\label{eq:vr}
\end{equation}

We use this model to esimate individuals' VR in real-time using heart rate data from a fitness watch or heart-rate monitor.

\subsection{Received Deposition Dose Model}
Various methods exist for calculating the deposition dose of particulates. For this study, a deterministic model outlined in \cite{PM_01} was used due to its low computational cost, which facilitated real-time calculation. The model estimates the proportion of particles inhaled per breath (IF, Equation \ref{eq:if}), then the proportion of those particles which are deposited in the lung (DF, Equation \ref{eq:df}). The PM2.5 mass-median diameter (MMD) for primary and secondary vehicle emissions at street level is approximated as constant and was estimated using data presented in \cite{PM_01}.

\begin{equation}
    \IF = 1 - 0.5 \left(1 - \frac{1}{1 + 0.00076 \text{ MMD}^{2.8}} \right) \\
    \label{eq:if}
\end{equation}

\begin{equation}
    \begin{aligned}
    \DF = \IF \bigg( 0.0587 + &\frac{0.911}{1 + e^{4.77 + 1.485 \ln(\text{MMD})}} +\\
    &\mspace{60mu} \frac{0.943}{1 + e^{0.508 - 2.58 \ln(\text{MMD})}} \bigg)
    \end{aligned}
\label{eq:df}
\end{equation}

The deposition fraction is then combined with the ventilation rate, duration of exposure $t$, and level of exposure $e$ to estimate RDD

\begin{equation}
    \RDD = \VR \cdot \DF \cdot t \cdot e
\label{eq:rdd}
\end{equation}

\section{Respiratory-Aware Routing}

As a preliminary evaluation of the model, we develop a simulated commuting system, where cyclists interact with a fixed map to travel between an origin and destination. We note that the method is not limited to cyclists, however, and future work will expand its application to large-scale mobility simulations such as SUMO \cite{SUMO}.

\subsection{Simulation Environment}
We model the system of roads and cycle lanes that cyclists traverse during a commute as a travel graph, where junctions are modelled as vertices and road segments as edges. This graph is randomly initialised with edge lengths, pollution exposures, and grades to simulate an urban or sub-urban environment. We further model the cyclists' behaviour over this map as an MDP, where cyclists can act in a given state to move to another state.

The simulation is turn-based, with cyclist actions chosen based on a travel-time priority queue to model travel duration along each edge. Congestion is modelled through limiting travel velocity to the leading cyclist when an edge is being traversed by two or more cyclists.

\subsection{Cyclist Model}
Given the simulated nature of this experiment, real-time heart rate data were approximated for cyclists by estimating the instantaneous power requirement of each action, and using a first-order differential model for heart rate.

The power required to traverse a given road segment at velocity $v$ was calculated from first principles as shown in Equations \ref{eq:power} and \ref{eq:sub_power}, with parameter estimations for commuter bikes taken from \cite{O_05} and \cite{O_07}.
\begin{equation}
    P = \frac{ P_{gradient} + P_{friction} + P_{air}}{\eta_{mech}}
\label{eq:power}
\end{equation}

\begin{equation}
    \begin{aligned}
    P_{gradient} &= m g \frac{\Delta h}{l} v\\
    P_{friction} &= C_r  m  g v\\
    P_{air} &= \frac{1}{2} C_d \rho A v^3\\
    \end{aligned}
\label{eq:sub_power}
\end{equation}

Cardiovascular drift is accounted for through increasing the perceived power output $P(t)$ in proportion with the energy expenditure since the start of the activity, as proposed in \cite{o_06}.

\begin{equation}
    \begin{aligned}
    P(t) = P(t) + k_f \sum_{t_0}^{t} P(t)
    \end{aligned}
\label{eq:power_history}
\end{equation}

The simulated cyclist's heart rate was then estimated based on this perceived power output using the first-order relationship given in Equation \ref{eq:hr}.

\begin{equation}
    HR(t) = 
    \begin{cases}
    \text{\bf IF } {\bf HR(t) < HR_{max}}\\ 
        \HR_{SS}\big(P(t)\big) + \Big(\HR_{SS}\big(P(t)\big) - \HR\left(t_0\right)\Big) e^{-\frac{t}{\tau_r}} \\ \\
    \text{\bf ELSE } \HR_{max} 
    \end{cases}\\
\label{eq:hr}
\end{equation}
\begin{equation*}
    \text{\bf where \;} \HR_{SS}\big(P(t)\big) = \HR(t_0) + c \cdot P(t)
\end{equation*}

\subsection{Fitness Modelling}
In order to model a representative population, commuting cyclists were discretized into three sub-populations according to endurance fitness. Physiological parameter ranges were estimated through extrapolation of the ranges provided in \cite{o_06} using data from the 2020 National Travel Survey \cite{NTS_2020}.

For each cyclist, physiological parameters were sampled from a Gaussian distribution parameterised by the mean and standard deviations given in Table \ref{tab:fitparams}. The sex of each cyclist was randomly chosen.

\begin{table}[h]
\caption{Sampled distributions for each fitness strata, mean (standard deviation). * mass denotes the mass of the cyclist and bicycle.}
\centering
\begin{tabular}{l | c c c}
\hline
\hline
 & Less Fit & ... & Fitter\\
\hline
Mass* (kg) & 95(10) & 85(10) & 75(5)\\
$HR_0$ (bpm) & 80(5) & 70(5) & 60(5)\\
$\tau_r$ (s) & 30(2) & 26(2) & 22(2)\\
$HR_{max}$ (bpm) & 180(5) & 180(5) & 190(5)\\
$c$ (bpm / W) & 1e-5(5e-6) & 3e-5(5e-6) & 6e-5(5e-6)\\
$k_f$ (W / J) & 0.45(0.02) & 0.3(0.05) & 0.15(0.02)\\
\hline
\end{tabular}
\label{tab:fitparams}
\end{table}

\subsection{Reinforcement Learning}
The simulation environment is expanded with observations $s(t)$, actions $a(t)$, and rewards $R(t)$ for each agent:
\begin{equation}
    s(t) = [x(t), g, c(t), p(t), h(t)]
\end{equation}
where $x(t)$ is the one-hot encoded position of the agent, $g$ is the one-hot encoded goal state, $c(t)$ is a vector encoding the cyclist's estimated physiological parameters, and $p(t)$ and $h(t)$ are pollution- and grade-weighted adjacency matrices.

The action space is made up of the nodes in the graph, with the action corresponding to a traversal of the edge $\left(x(t), a(t)\right)$. Invalid actions, where this edge does not exist in the graph, are encoded as zeros in the observation matrices.



Reward is assigned according to Equation \ref{eq:reward} where $\alpha$ and $\beta$ are tuning parameters set to $-50$ and $1$ respectively. Agents therefore receive a small negative reward for valid actions that do not reach the goal, a large negative reward for invalid actions, and a positive reward for reaching the goal.

\begin{equation}
    \begin{aligned}
    R(s, a) =& \alpha \RDD(s, a) + \beta \isgoal(s, a)\\
    \end{aligned}
\label{eq:reward}
\end{equation}

The active agent at time $t$ is selected using a priority queue, where agents are assigned priority inversely proportional to the total duration of their journey, such that agents who have travelled for the least duration (i.e. taken the fewest steps or the fastest path) have the highest priority. Valid actions result in mutation of this queue according to the new journey duration, while invalid actions result in no movement and are penalised through large negative reward.

Training is conducted using PPO for multi-agent environments implemented in Ray's RLLib \cite{rllib} and using the PettingZoo framework \cite{terry2021pettingzoo} on a MacBook Pro (M1) and Python 3.9.11, for up to 5000 iterations. Code to replicate the experiments is available on the author's GitHub\footnote{\url{https://github.com/Langbridge/RL_RAR}}.

\subsection{Experiments}
To benchmark the behaviour of the system, a simple dynamic search tree was utilised where each node contains information regarding the cyclist's current state, including heart rate and power history. This approach allows the dynamic nature of the routing problem to be modelled and solved exactly, but is not suitable for large problems due to significant memory overhead.

\begin{figure}[h!]
\centering 
\subfloat[Terrain map.]{\includegraphics[width=0.45\columnwidth]{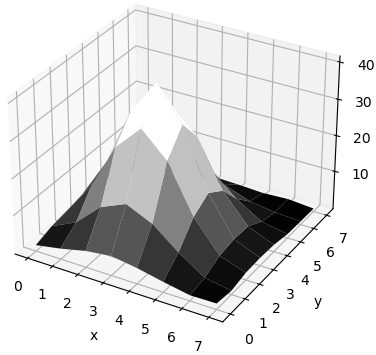}} \hspace{0.2pt}
\subfloat[Pollution map.]{\includegraphics[width=0.45\columnwidth]{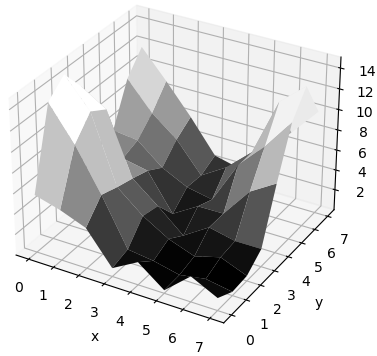}} \\
\caption{8x8 grid map with central hill and polluted areas.}
\label{fig:8x8map}
\end{figure}

A simple 8x8 map with a large hill at the centre of the grid and areas of high pollution centred on three of the map corners as shown in Figure \ref{fig:8x8map} was used to benchmark. Each node is connected to those directly adjacent with an edge length sampled from the distribution $L\sim \mathcal{N}(500, 20^2)$. The pollution-optimal paths from (0,0) to (7,7) were then calculated for a fitter and less fit cyclist sampled from the distributions outlined in Table \ref{tab:fitparams}.

\begin{figure}[h!]
\centering 
\subfloat[Terrain map.]{\includegraphics[width=0.45\columnwidth]{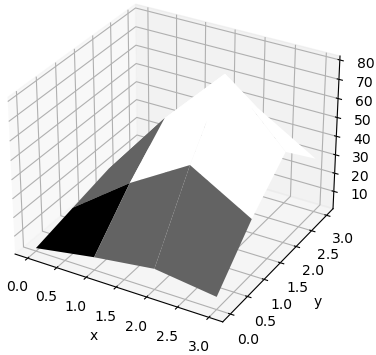}} \hspace{0.2pt}
\subfloat[Pollution map.]{\includegraphics[width=0.45\columnwidth]{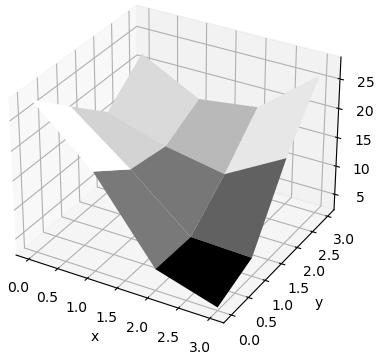}} \\
\caption{Half-size version of the 8x8 grid map with central hill and polluted areas.}
\label{fig:4x4map}
\end{figure}

For the initial RL experiments, the half-scale map in Figure \ref{fig:4x4map} was produced to accelerate the learning for the proof-of-concept demonstration. This allows the convergence of the policy's pollution optimality to be evaluated for each checkpoint while minimising the computational expense of repeated calls to the dynamic search. The policy was trained over a fixed map with ten random agents for 5,000 iterations, to encourage the policy to generalise and be robust to finding optima for different cyclist parameters.

Finally, to demonstrate the scalability of the approach to larger maps, another policy was trained on the 8x8 map. The policy was trained with ten random agents for 5,000 iterations. For this larger map, the goal is constrained to be within a $k$-hop neighbourhood of the origin, with $k$ set according to:
\begin{equation}
    k = \max \left\{ 1, \left\lfloor 2+\frac{i-500}{500}\right\rfloor \right\}
\end{equation}
where $i$ is the training iteration. This goal scheduling allows the policy to learn to navigate to the goal on a larger map without having to mutate the map itself during training.


\section{Results}

\subsection{Individualised Routing}
\begin{figure}[h!]
\centering 
\subfloat[Optimal routes over terrain map.]{\includegraphics[width=0.45\columnwidth]{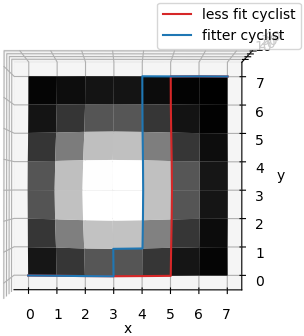}} \hspace{0.2pt}
\subfloat[Optimal routes over pollution map.]{\includegraphics[width=0.45\columnwidth]{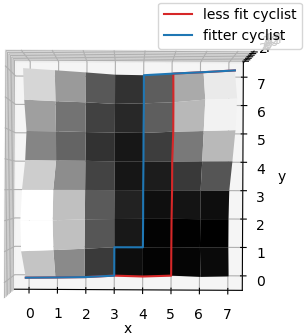}} \\
\caption{Optimal routes for a fitter and less fit cyclist over the 8x8 map.}
\label{fig:brute-routes}
\end{figure}

As shown in Figure \ref{fig:brute-routes}, the optimal routes for the two cyclists differ even over this relatively small map. It's clear from Figure \ref{fig:brute-routes}a that the pollution dose received by less fit cyclist is more significantly impacted by the power required to ascend the hill than the fitter cyclist, which takes the minimal exposure route over the hill. The differences in the route characteristics between cyclists, and the pollution dose as compared to the shortest path, are summarised in Table \ref{tab:brute-route-eval}.

\begin{table}[h]
\caption{Comparison of optimal route characteristics between a random fitter and less fit cyclist. The shortest path's RDD is included for comparison.}
\centering
\begin{tabular}{l | c c | c}
\hline
\hline
 & Less Fit & Fitter & \% Difference\\
\hline
Mean Exposure ($\mu g m^{-3}$) & 3.82 & 3.45 & -9.67\\
Mean Elevation ($m$) & 4.82 & 9.24 & 91.9\\
RDD ($\mu g$) & 0.203 & 0.0821 & -59.5\\
\hline
Shortest Path RDD ($\mu g$) & 0.690 & 0.256 & -62.8\\
\hline
\end{tabular}
\label{tab:brute-route-eval}
\end{table}

\subsection{Policy Routes}
Table \ref{tab:4x4-policy-eval} compares the RDD for three possible routes: the policy path, the shortest path, and the pollution-optimal path. It's clear that the policy is able to learn to approximate the pollution risk model well over this small map, achieving optimality higher than 98\% for random cyclists with various levels of fitness not seen during training.

\begin{table}[h]
\caption{Comparison of pollution dose for a random fitter and less fit cyclist when routed using the learned policy, brute force, and the shortest path over the 4x4 map.}
\centering
\begin{tabular}{l | c c}
\hline
\hline
 & Less Fit & Fitter\\
\hline
Policy RDD ($\mu g$) & 0.834 & 0.168\\
\textit{Optimality} & \textit{0.982} & \textit{1.000}\\
\hline
Shortest Path RDD ($\mu g$) & 1.380 & 0.275\\
\textit{Optimality} & \textit{0.593} & \textit{0.611}\\
\hline
Optimal RDD ($\mu g$) & 0.819 & 0.168\\
\hline
\end{tabular}
\label{tab:4x4-policy-eval}
\end{table}

\subsection{Scalability}
\begin{figure}[h!]
\centering 
\subfloat[Policy routes over terrain map.]{\includegraphics[width=0.45\columnwidth]{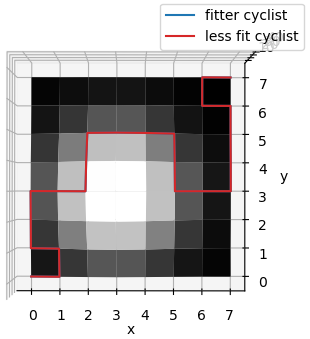}} \hspace{0.2pt}
\subfloat[Policy routes over pollution map.]{\includegraphics[width=0.45\columnwidth]{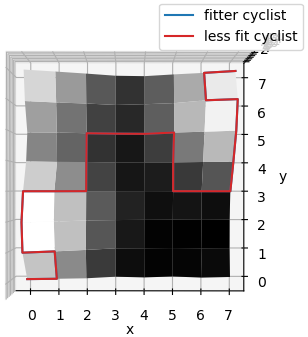}} \\
\caption{Policy routes for a fitter and less fit cyclist over the 8x8 map.}
\label{fig:8x8-policy-eval}
\end{figure}

The behaviour of the policy trained on the 8x8 map is less definitive, however. While valid paths are consistently produced after 5,000 iterations, the paths are clearly non-optimal as shown in Figure \ref{fig:8x8-policy-eval} and Table \ref{tab:goal-schedule-eval}. Further, the policy paths are identical for both levels of fitness, suggesting that the pollution risk model is not well approximated by the policy for this more complex environment.

\begin{table}[h]
\caption{Comparison of pollution dose for a random fitter and less fit cyclist when routed using the 8x8 learned policy, brute force, and the shortest path. Note that the given RDD differs from Table \ref{tab:brute-route-eval} due to the random sampling of cyclists.}
\centering
\begin{tabular}{l | c c}
\hline
\hline
 & Less Fit & Fitter\\
\hline
Policy RDD ($\mu g$) & 1.096 & 0.278\\
\textit{Optimality} & \textit{0.214} & \textit{0.268}\\
\hline
Shortest Path RDD ($\mu g$) & 0.536 & 0.166\\
\textit{Optimality} & \textit{0.437} & \textit{0.449}\\
\hline
Optimal RDD ($\mu g$)& 0.234 & 0.0745\\
\hline
\end{tabular}
\label{tab:goal-schedule-eval}
\end{table}

\section{Discussion}
It's clear from the results presented in this work that the impact of physiological fitness on pollution risk is significant, even when pollution risk is optimised for. Further, with more than 40\% reduction in pollution risk for the pollution-optimal path as compared to the shortest path, the benefits of respiratory-aware routing are clear even in this trivial case. In environments with greater terrain or pollution variation, as is likely in urban areas, this benefit is likely to be even more pronounced.

We demonstrate that a reinforcement learning policy can learn to minimise this risk for a diverse population of cyclists across a small map, producing near-optimal routes for unseen origin-destination pairs and cyclists. This exhibits the power of these models to generalise to diverse scenarios with inference time significantly faster than the dynamic search tree. This illustrates the value of RL approaches to approximate this highly dynamic model for real-time routing applications.

However, the results from the policy trained on the 8x8 map are less clear. While goal-scheduling enables the policy to produce valid paths to the goal in only 5,000 iterations, the paths produced are far from optimal on what remains a trivially-sized map. Despite these shortcomings, the policy avoids the apex of the hill, which corresponds to our intuitive understanding of minimising VR to minimise pollution inhaled. Performance would likely be improved with longer training on dedicated hardware, or through the addition of some heuristic term to the reward function to encourage movement towards the goal.

There is significant opportunity to expand this proof-of-concept work. By expanding the action space and implementing variable velocity, the effects of congestion on path optimality could be explored. The RL architecture could then be expanded to introduce hierarchy, where a high-level controller selects policies to balance greedily minimising pollution risk and ensuring that pollution risk is allocated fairly amongst cyclists. This latter point, on minimising pollution risk globally for multiple agents, will be treated with particular care due to the ethical implication of suggesting a route to a specific user to minimise the global pollution risk at the potential expense of the individual pollution risk.

Further, the graph structure of the environment facilitates the expansion of the map to more complex systems with communities and highly congested connecting edges similar to the network structure city centres such as Dublin which are built across a central river. This graph structure also creates the potential for more expressive actor networks, such as through the implementation of Graph Neural Network (GNN) layers to reason about the graph structure explicitly.

Finally, we intend to implement our controller on more realistic large-scale simulators, such as SUMO \cite{behrisch2011sumo}, in order to validate our results on a state-of-the-art environment. Further, we are currently collecting real-time pollution and heart rate data from urban cyclists to evaluate our proposed model in the real world.

\section{Conclusion}
In this work, a proof-of-concept system for respiratory-aware routing was presented. The work highlights the remarkable variation of pollution risk between different cyclists travelling between the same origin and destination. In particular, individuals' pollution inhalation is strongly dependent on their fitness, with pollution-optimal routes varying even on small-scale maps.

The potential of reinforcement learning to extend the method to real-time routing was also explored. We show that policies can achieve more than 98\% optimality compared to a dynamic programming approach, however we recognise the complexity of tuning and training policies for larger environments. Future work will attempt to overcome this to achieve scalability to larger, more complex systems, including integration with SUMO.


\bibliographystyle{ieeetr}
\bibliography{ref.bib}



\end{document}